\documentclass[twocolumn,trackchanges]{aastex63}
\submitjournal{ApJ}

\shorttitle{Atmospheric escape}
\shortauthors{Mitani et al.}
\graphicspath{{./}{figures/}}
\newcommand{\eV}{{\rm \, eV}}
\newcommand{\reffig}[1]{Figure~\ref{#1}}
\newcommand{\refsec}[1]{Section~\ref{#1}}
\usepackage{mhchem}
\usepackage{here}

\begin{document}


\title{Atmospheric Escape of Close-in Giants around Hot Stars: Far-Ultraviolet Radiation and Photoelectric Heating Effect}

\correspondingauthor{Hiroto Mitani}
\email{hiroto.mitani@phys.s.u-tokyo.ac.jp}

\author{Hiroto Mitani}
\affiliation{Department of Physics, School of Science, The University of Tokyo, 7-3-1 Hongo, Bunkyo, Tokyo 113-0033}

\author{Riouhei Nakatani}
\affiliation{RIKEN Cluster for Pioneering Research, 2-1 Hirosawa, Wako, Saitama 351-0198, Japan}

\author{Naoki Yoshida}
\affiliation{Department of Physics, School of Science, The University of Tokyo, 7-3-1 Hongo, Bunkyo, Tokyo 113-0033}
\affiliation{Kavli Institute for the Physics and Mathematics of the Universe (WPI), UT Institutes for Advanced Study, The University of Tokyo, Kashiwa, Chiba 277-8583, Japan}
\affiliation{Research Center for the Early Universe, School of Science, The University of Tokyo, 7-3-1 Hongo, Bunkyo, Tokyo 113-0033}


\begin{abstract}
Atmospheric escape is an important process that controls the long-term evolution of close-in planets. We perform radiation hydrodynamics simulations of photo-evaporation of exoplanets' atmospheres to study the effect of photoelectric heating by far-ultraviolet (FUV) radiation.
Specifically, we consider a close-in hot Jupiter around a hot A-star. 
Hot main-sequence stars emit not only extreme ultraviolet radiation but also FUV radiation, and thus can drive strong atmospheric escape by photoelectric heating.
We show that the planetary atmosphere escapes at a rate as large as $\dot{M}\sim10^{14}\,\mathrm{g}~{\rm sec}^{-1}$ if the atmosphere contains a small amount of dust grains with the level of ten percent of the local interstellar medium.
Close-in planets around hot stars can lose a significant fraction of the atmosphere during the long-term evolution. 
We quantify the amount of dust necessary for causing photoevaporation. The dust-to-gas mass ratio of $10^{-4}$ is sufficient to drive stronger atmospheric escape by FUV photoelectric heating than in the case with only extreme ultraviolet radiation.
We also explore the metallicity dependence of the 
FUV-driven escape. The mass-loss rate increases with increasing the atmosphere's metallicity because of the enhanced photoelectric heating, but the stellar FUV flux decreases with increasing stellar metallicity. 
We derive an accurate estimate for the mass-loss rate as a function of FUV flux and metallicity, and of the planet's characteristics. 
The FUV driven atmospheric escape may be a key process to understand and explain the so-called sub-Jovian desert.

\end{abstract}



\section{Introduction} \label{sec:intro}
Hot Jupiters are gas giant planets orbiting close to the host star. Intense radiation from the star can heat the planets' atmosphere, and photo-evaporation, or the so-called atmospheric escape, can be driven by heating associated with ultraviolet or X-ray photoionization \citep{Yelle_2004, Tian_2005, Murray-Clay_2009, Owen_2012}.
Spectroscopic observations during transit have revealed atmospheric escape from exoplanets such as HD209458b \citep{Vidal-Madjar_2003} and GJ 436b \citep{Ehrenreich_2015}. 
It is thought that atmospheric escape is one of the key processes that determine the long-term evolution of a gaseous planet.

Interestingly, recent statistical studies of exoplanet populations suggest absence of intermediate-mass ($0.02M_\mathrm{J}<M_p<0.8M_\mathrm{J}$) and short-period ($P<3$ d) planets, which is called ``sub-Jovian desert'' \citep{Szabo_2011,Mazeh_2016}. 
Similar paucity is also reported for close-in planets with radius of $\sim1.8R_{\oplus}$ \citep{Fulton_2017}. 
\citet{Szabo_2019} suggests that the boundary of the sub-Jovian desert depends on the metallicity and the effective temperature of the host star. 
\citet{Allan_2019,Owen_2017} argue that these trends can be explained by atmospheric escape driven by UV/X-ray radiation from the host star. 
If the radiation-driven atmospheric escape is the dominant physical process that can explain these observed features, it is important to explore theoretically the possible metallicity and stellar temperature dependence of the atmospheric escape.

Previous theoretical studies focus on the effect of photoionization heating by extreme-ultraviolet (EUV; $13.6 \eV < h\nu < 100\eV$) photons and X-rays. 
There are also many studies that consider atmospheric escape from close-in planets around sun-like stars \citep{Murray-Clay_2009, Koskinen_2013, Owen_2012,Owen_2014,Tripathi_2015}. 
Recent observations have discovered several hot Jupiters around hot A-stars (KELT-9b; \citep{Gaudi_2017}, KELT-20b; \citep{Lund_2017}). 
It is important to notice that a hot A-star emits a copious amount of far-UV (FUV; $h\nu <13.6\eV$) photons that can also cause photoelectric heating, while it emits relatively weak EUV radiation because of the lack of surface convection \citep{Fossati_2018}. 
Unfortunately, the effect of FUV radiation has not been
studied in detail in the context of exoplanet atmospheric escape. 
FUV photoelectric heating of dust grains and polycyclic aromatic hydrocarbons (PAHs) is known to be a major heating mechanism in star-forming gas clouds and in the interstellar medium \citep[e.g.,][]{Watson_1972, Jura_1976, Bakes_1994}. 
We argue that the same heating process can be important in, for example, close-in dusty planets around hot stars that have been discovered \citep[e.g.][]{Pont_2013, Nikolov_2015}.  
It thus appears timely and important to explore a wide parameter space of the stellar FUV flux and dust-to-gas mass ratio, in order to identify physical conditions where FUV photoelectric heating can be effective to drive the atmospheric escape. 

In the present paper, we perform radiation hydrodynamics simulations of atmospheric escape from a close-in gas giant irradiated by a hot star. 
Our simulations include the radiative transfer of EUV and FUV photons and non-equilibrium chemistry and thus allow us to examine the thermal evolution of the planet atmosphere and the effect of FUV photoelectric heating. We run a set of simulations with systematically varying the metallicity 
in the planetary atmosphere. We also investigate the stellar effective temperature dependence of the atmospheric escape rate.
The rest of the paper is organized as follows. In \refsec{sec:Models}, we present the models of our simulations. In \refsec{sec:Results}, we show the results of our simulations. In \refsec{sec:Discussion} and \refsec{sec:summary}, we discuss the mass loss due to FUV heating and give a summary, respectively.

\section{Numerical Model}\label{sec:Models}
We follow the dynamical evolution of a close-in, sub-Jovian-mass planet's atmosphere irradiated by both EUV and FUV radiation from the hot host A-star.
We use the hydrodynamics simulation code PLUTO \citep{Mignone_2007} suitably modified for our study \citep{Nakatani_2018, Nakatani_2018b}.

Assuming symmetry around an axis parallel to the EUV/FUV field, we adopt 2D cylindrical polar coordinates and solve the following hydrodynamic equations coupled with radiative transfer:
\begin{eqnarray}
\frac{\partial\rho}{\partial t}+\nabla\cdot\rho\vec{v} &=& 0\\
\frac{\partial\rho v_R}{\partial t} +\nabla\cdot(\rho v_R\vec{v}) &=& -\frac{\partial P}{\partial R}-\rho\frac{\partial\Psi}{\partial R}\\
\frac{\partial\rho v_z}{\partial t} +\nabla\cdot(\rho v_z\vec{v}) &=& -\frac{\partial P}{\partial z}-\rho\frac{\partial\Psi}{\partial z}\\
\frac{\partial\rho E}{\partial t}+\nabla\cdot(\rho H\vec{v})&=&-\rho\vec{v}\nabla\Phi+\rho(\Gamma-\Lambda)
\end{eqnarray}
where $\rho, \vec{v}, P, \Psi$ are gas density, velocity, pressure, and gravitational potential of the star and planet including the centrifugal force term. We configure plane-parallel stellar radiation coming from negative $z$ direction. We neglect the self-gravity of the atmosphere because the mass of the upper atmosphere is much smaller than the planet mass. The relevant heating ($\Gamma$) and the cooling ($\Lambda$) processes are described in \refsec{sec:heating_cooling}. We also follow 
non-equilibrium chemistry 
\begin{eqnarray}
\frac{\partial n_H y_i}{\partial t}+\nabla\cdot(n_H y_i \vec{v})=n_H R_i
\end{eqnarray}
where $y_i = n_i/n_H$ and $R_i$ represent the abundance and the reaction rate, respectively. The incorporated chemical species are \ce{H, H+, H_2, CO, C+, O, e-}.

Our code incorporates ray-tracing radiative transfer of EUV and FUV photons as in \citet{Nakatani_2018, Nakatani_2019}. We describe further details of our simulations in \refsec{sec:heating_cooling}.

The fiducial model parameters are listed in Table~\ref{tab:fid_params}. The values are chosen for the atmospheric escape from planets around hot A-stars.
Our model is also aimed at addressing possible physical causes of the sub-Jovian desert. 
We calculate the stellar FUV flux from the spectral model 
of a 10000\,K star \citep{Husser_2013} by integrating the flux in the range of 6\,eV to 13.6\,eV. 
For EUV, we use the model spectrum of \citet{Fossati_2018} for a 10000\,K star.
We perform additional simulations in order to study the effect of the stellar spectrum and of the metallicity. The model variation is introduced in \refsec{sec:Results}.

\begin{table}[h]
    \centering
        \caption{Model parameters in the Fiducial Run}
    \begin{tabular}{ll}\hline\hline
        Stellar parameters & \\
        Stellar Mass $M_*$ & 3$\mathrm{M}_{\odot}$ \\  
        Stellar Radius $R_*$ & 1.6$\mathrm{R}_{\odot}$ \\
        Stellar EUV photon emission rate &  $4.4\times10^{38}$ photons/s       \\
        Stellar FUV luminosity &   $1.5\times10^{34}$ erg/s     \\ \hline
        Planetary parameters & \\
        Planet Mass $M_p$& 0.3 $\mathrm{M}_\mathrm{J}$\\ 
        Planet Radius $R_p$& 1$\mathrm{R}_\mathrm{J}$\\
        Semi-major axis $a$ & 0.045\, AU \\ 
        Metallicity $Z$ & $0.1Z_{\odot}$\\
        Dust-to-gas mass ratio & 0.001\\ \hline
    \end{tabular}
    \label{tab:fid_params}
\end{table}

\begin{figure*}[h]
\includegraphics[width=18cm]{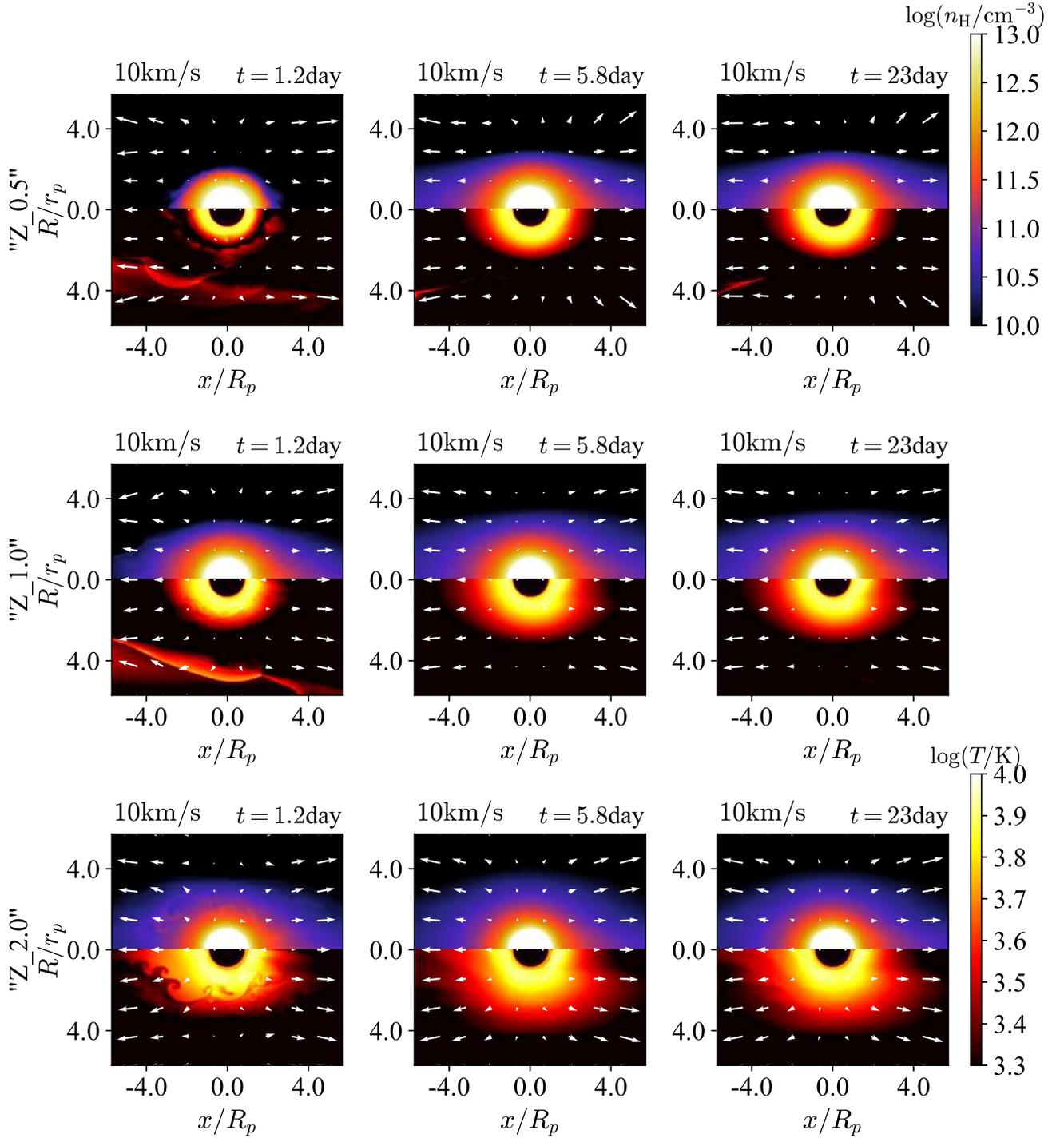}
\caption{Structure of the planetary atmosphere in simulations with different metallicities: $Z=0.5\times\mathrm{fiducial}$ (Top rows), $Z=\mathrm{fiducial}$ (Middle rows), $Z=2\times\mathrm{fiducial}$ (Bottom rows) at $t=1.2\mathrm{days}$ (left), $t=5.8\mathrm{days}$ (middle), and $t=23\mathrm{days}$ (right). 
The host star is located to the left. In each panel, the upper half shows the density profile and the lower half shows the temperature profile, and the arrows show the gas velocity field. }
\label{fig:metallicity_structure}
\end{figure*}

\subsection{Heating  processes}\label{sec:heating_cooling}
Our simulations include the photoionization heating by EUV photons and 
photoelectric heating by FUV photons. 

We adopt the photoelectric heating rate of \citet{Bakes_1994}:
\begin{eqnarray}
\Gamma_{\mathrm{pe}} &=& 10^{-24}\,\epsilon \, G_\mathrm{FUV} \,n_{\mathrm{H}} Z/Z_{\odot}\ \mathrm{ergs\ s^{-1}\ cm^{-3}}\\
\epsilon &=& \frac{4.87\times10^{-2}}{1+4\times10^{-3}\,(G_{\mathrm{FUV}}\sqrt{T}/n_{e})^{0.73}} \nonumber\\
&+& \frac{3.65\times10^{-2}(T/10^{4}\mathrm{ \ K})^{0.7}}{1+2\times10^{-4}\, G_{\mathrm{FUV}}\sqrt{T}/n_{e}}
\label{eq:pe_rate}
\end{eqnarray}
where $\epsilon$ is the heating efficiency that depends on the gas temperature $T$, electron density $n_e$ and the normalized 
local FUV flux $G_\mathrm{FUV}$, and $n_{\mathrm{H}}$ is the hydrogen nuclei number density. The FUV flux is computed as $G_\mathrm{FUV} = F_0 \exp(-1.8 A_\mathrm{V}) /(1.6\times10^{-3}\mathrm{\,erg}\mathrm{\,s}^{-1}\mathrm{\,cm}^{-2})$, where $A_\mathrm{V}$ is the visual extinction. 

The adopted heating efficiency in Equation (\ref{eq:pe_rate}) implicitly assumes ISM-like carbonaceous grains \citep{Mathis_1977} with a dust-to-gas mass ratio of $0.01$ for $Z = 1\, Z_\odot$. 
Although little is known about the nature of dust in exoplanet atmosphere, our choice here is motivated by the fact that Titan's upper atmosphere is known to have a PAH abundance similar to that of the ISM \citep{Lopez-Puertas_2013}. Interestingly, disequilibrium chemistry models of \citet{Moses_2013} show that the graphite clouds can exist in the hot atmosphere if the C/O ratio is sufficiently large. In such a planetary atmosphere, FUV photoelectric heating can be an important 
process, and thus it is worth examining the impact on atmospheric escape.

Dust grains are essential for FUV photoelectric heating. Theoretical chemical calculations of hydrostatic atmosphere \citep{Morley_2012, Lavvas_2017} show that {\it in situ} dust formation in the upper atmosphere ($\sim 10^{-6}-10^{-9}$ bar) is inefficient. However, 
small dust grains can be transported from the lower to upper atmosphere when strong hydrodynamical escape is occuring \citep{Wang_2018,Wang_2019}.
The drag force exerted on a moving grain in a gas with temperature $T$ and number density $n$ is given by 
\begin{equation}
    F_{\mathrm{drag}}\approx2\pi a^2 kTn\times\frac{8s}{3\sqrt{\pi}}\left(1+\frac{9\pi}{64}s^2\right),
\end{equation}
 where $a$ is the grain radius and $s = \sqrt{\mu v^2/ 2kT}$ with $\mu$ is the mean molecular mass \citep{Baines_1965,Draine_2011}. Equating the drag force to the gravitational force, we derive
\begin{equation}
    \begin{split}
        \frac{GM_pM_d}{r^2}\times\frac{3\sqrt{\pi}}{16\pi a^2kTn}&=s\left(1+\frac{9\pi}{64}s^2\right)\\
        &>s=\sqrt{\mu v^2/ 2kT}
    \end{split}
\end{equation}
Hence, if the atmosphere is escaping at a speed greater than
\begin{equation}
    \begin{split}
    v_{\rm c} &= 10.1 \mathrm{\,m\,s}^{-1} \times \left(\frac{M_p}{M_J}\right)\left(\frac{r}{R_J}\right)^{-2}\left(\frac{a}{10\mathrm{\AA}}\right)\\
    \times&\left(\frac{\rho_d}{3\mathrm{\,g\,cm}^{-3}}\right)\left(\frac{n}{10^{12}\mathrm{\,cm}^{^3}}\right)^{-1}\left(\frac{T}{10^3\mathrm{K}}\right)^{-1/2}\left(\frac{\mu}{m_p}\right)^{-1/2},
    \end{split}
\end{equation}
the dust grains can be transported to upper layers by the drag force. Note that the velocity condition is weaker for smaller dust grains.
Defining the dust mass as $M_d=4/3\pi a^3\rho_d$, the condition is rewritten 
in terms of the photoevaporation rate as
\begin{equation}
    \begin{split}
    \dot{M}&>1.0\times10^{12}\mathrm{\,g\,s}^{-1}\times\left(\frac{M_p}{M_J}\right)\left(\frac{a}{10\mathrm{\AA}}\right)\\
    &\times\left(\frac{T}{10^3\mathrm{K}}\right)^{-1/2}\left(\frac{\mu}{m_p}\right)^{1/2}.
    \end{split}
\end{equation}
Note that the contribution from EUV-driven outflows
should be included when considering the above condition.
When this condition is met, small dust grains can be advected upward together with
an escaping gas, and thus can be present in the upper atmosphere if they can survive.
Recent detailed calculations with haze coagulation, sedimentation, diffusion and advection by planetary outflows show that small hazes can exist in a very low pressure ($<10^{-8}$ bar) region in puffy planets \citep{Gao_2020}. Therefore, we suggest that there can exist at least a small amount of dust grains in the atmosphere of gas giant planets such as those in the sub-Jupiter desert.

Cloud condensation can occur in layers with temperatures below 2000K \citep{Mbarek_2016}. In very hot exoplanet atmosphere, it is unlikely for clouds and hazes to form \citep{Helling_2019}, but a recent experimental study on photochemistry revealed that a solid organic could condensate under the influence of UV radiation \citep{Fleury_2019}. Planets around metal-rich host star may also have thick clouds 
containing Al and Ti \citep{Wakeford_2017}. 
All these recent findings further motivate us to study the thermochemical effect of small grains on atmospheric escape.

The dust-to-gas mass ratio in hot Jupiters is suggested to be $\sim0.001$ \citep{Helling_2016,Helling_2019,Woitke_2020}. We assume that the model planetary atmosphere contains ISM-like heavy elements and dust with $Z=0.1Z_{\odot}$ as our fiducial model, but
we treat the dust-to-gas mass ratio (metallicity) as a model parameter and explore a wide parameter space to identify the conditions
where the photoelectric heating is effective to drive atmospheric escape.

We assume a uniform dust-to-gas mass ratio over the computational domain for simplicity. This assumption is valid when photo-evaporative flows replenish small dust grains from the flow base into the upper atmosphere \citep{Wang_2018}. 
We will discuss the destruction/sublimation of dust/PAH and the influence on our results in \refsec{sec:dustpah}.

\subsection{Cooling processes}
The major cooling processes incorporated in our simulations are recombination cooling of \ion{H}{2} \citep{Spitzer_1978}, Ly$\alpha$ cooling of \ion{H}{1} \citep{Anninos_1997}, fine structure line cooling of \ion{O}{1} and \ion{C}{2} \citep{Hollenbach_1989, Osterbrock_1989, Santoro_2006} 
and molecular line cooling of \ce{H2} and \ce{CO} \citep{Galli_1998,Omukai_2010}.  

We note that adiabatic cooling is the most important process to reduce the internal energy of the gas
\begin{eqnarray}
\Lambda_{\rm adi} = -P\frac{d}{dt}\frac{1}{\rho}, 
\end{eqnarray}
as we will see in \refsec{sec:Results}.

\subsection{Initial and boundary conditions}\label{sec:condition}
The computational domain 
is defined on a region with $R=[0,4]\times10^{10}\,\mathrm{cm}$ and $z=[-4,4]\times10^{10}\,\mathrm{cm}$. 
The domain is configured with the numbers of cells $(N_R, N_z) = (480,960)$.

The initial density profile of the atmosphere is given by a hydrostatic isothermal model
\begin{equation}
\rho(r) = \rho_p\exp\left[\frac{GM_p}{c_s^2}\left(\frac{1}{r}-\frac{1}{R_p}\right)\right],
\end{equation}
where $r$, $\rho_p$ and $c_\mathrm{s}$ are the radius measured from the planet center, the density at the surface of the planet ($r=R_p$), and the sound speed, respectively. 
The upper atmosphere of a hot Jupiter is likely to have a cool lower layer and a hot upper layer \citep{Murray-Clay_2009}. We set the initial temperature of the lower atmosphere ($r<1.1R_p$) to 2000\,K and that of the upper atmosphere ($r>1.1R_\mathrm{p}$) to 10000\,K. The pressure gradient is set to be continuous across the boundary between the two layers.  

We expect that the inner core region of the planet does not affect the structure of the upper atmosphere.
We thus do not follow the dynamical evolution of the inner region ($r<0.85R_\mathrm{p}$). 
The temperature and the density there are fixed throughout the simulation, 
and the outward velocity is set to $v_{\rm out} = 0$ so that there should be no inflow of material from the core.
The Mach number and the density contrast are empirically set to 2 and 1 at the outer boundary, respectively, to avoid spurious reflections and nonphysical accumulation of the escaping gas. By running several test runs, we have confirmed that this somewhat artificial boundary conditions do not significantly compromise our results. The mass-loss rate during the steady state increases by only 30\%, and there is little effect on the velocity field compared to the run in which we set a larger computational domain that covers the whole transonic region of the photo-evaporative flows with the conventional outflow boundary condition.
At the symmetry axis, we adopt conventional reflective axisymmetric boundary conditions.

We note that simulated atmospheric structure can be affected by the physical conditions of outflows at the outer boundary of the computational domain. Typically, the upper atmosphere shrinks and the temperature decreases as the outflow velocity increases \citep{Tian_2013}. We have run test simulations with varying the position of the outer boundary, to confirm that the outer boundary conditions do not affect our results. This can be understood by the fact that the photoevaporative flows have supersonic velocities at the outer boundary in all our simulations.

It is important to resolve the launching layer of photo-evaporative flows, especially for accurate estimation of the mass-loss rate. On the other hand, our simulations are computationally very expensive mainly because we follow non-equilibrium chemistry. 
The volume of the computational domain and the outer boundary conditions are chosen so that we can maximize the resolution in the regions of interest while saving computational cost.


\section{Results}\label{sec:Results}

\subsection{photo-evaporation rate}\label{sec:Fid}

\reffig{fig:metallicity_structure} shows the atmospheric structure in our
simulations. We first discuss our fiducial case shown in the middle row. 
The system reaches a quasi-steady state in $t\sim5$ days, which is about 50 times the crossing time. The gas temperature of the atmospheric surface is close to $10^4$\,K, being sufficiently high to drive atmospheric escape against the planetary gravity. The speed of the flow reaches $\sim10\,\mathrm{km/s}$ (close to $c_{\mathrm{s}})$ at around $r>4R_{\mathrm{p}}$.

\begin{figure}[h]
    \centering
    \includegraphics[width=9cm]{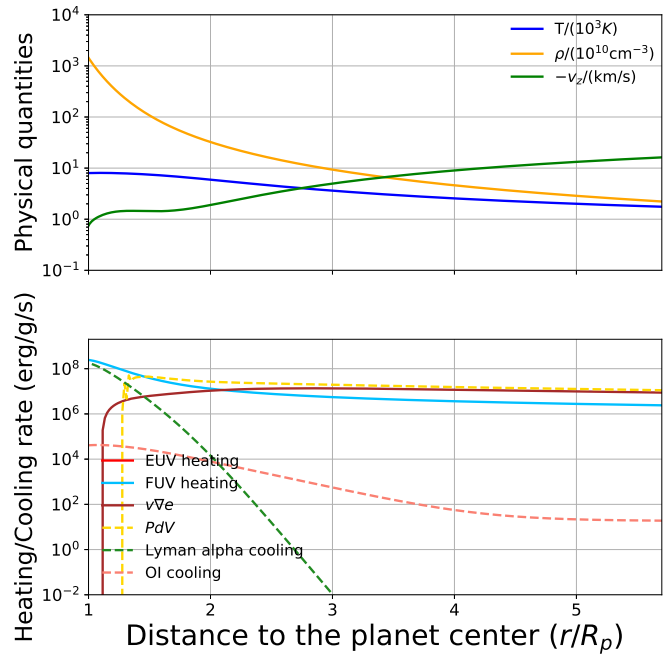}
    \caption{Time-averaged radial profiles of fiducial models at $R=0.01 R_p$ during $t=11.5$ d to $t = 23$ d. The fluid quantities ($T, \rho, -v_z$; Top panel) and the heating (solid line) and cooling (dashed line) rate (Bottom panel) are shown. EUV heating rate is much smaller than the other heating rate and framed out.}
    \label{fig:fiducial_radial}
\end{figure}

\reffig{fig:fiducial_radial} shows the time-averaged profiles of various quantities along the line defined at $R = 0.01 R_p$ and $z > 0$.
FUV photoelectric heating is the most efficient heating process. The energy advection $v\nabla e$ is important at $r>2R_{\mathrm{p}}$. In our simulations, the gravitational force by the central star exceeds the sum of the centrifugal force and the planet's gravity at $r > R_{\mathrm{Hill}}= a \sqrt[3]{M_p/3M_*}\sim 3R_{\mathrm{p}}$. 
Beyond $r \sim 3 R_{\mathrm{p}}$, the gas flows nearly
adiabatically under the influence of the host star's gravity. In this region, the flow is essentially isoentropic, and the pressure follows $p\propto T^\gamma/(\gamma-1)$, where $\gamma$ is the heat capacity ratio. It is interesting that EUV heating is negligible and is out of the frame in the lower panel of \reffig{fig:fiducial_radial}.
FUV reaches denser interior of the atmosphere, and the excited flows are optically thick to EUV. 

We find that the so-called $PdV$ cooling is the most efficient cooling process, and all the other cooling sources such as metal line cooling are sub-dominant. 
Hence the net cooling rate is independent of the gas metallicity. FUV heating balances the Lyman alpha cooling in the inner region.
The is true for all the other simulations performed in this study.

The abundance of heavy element and dust grains affects the heating efficiency in the planetary atmosphere, but gas-phase metal species does not significantly
contribute to the overall cooling rate.
In this sense, "metallicity" effectively corresponds to the dust-to-gas mass ratio in the present study.


We calculate the mass-loss rate $\dot{M}$ by integrating the mass flux across the 
square boundary defined by $|z|=2.9R_p,R=2.9R_p$. After the system reaches a quasi-steady state, 
the mass-loss rate remains
nearly constant at $\dot{M}\sim10^{14}\, \mathrm{g/s}$, which is much larger than typical EUV photo-evaporation rate of the order of  $\dot{M}\sim10^{9-11}\, \mathrm{g/s}$ \citep{Murray-Clay_2009, Owen_2012, Allan_2019}. 

We have run a test simulation where FUV heating is disabled. This run may be roughly corresponding to the case for a sun-like star. The mass-loss rate of this run is consistent with those of the previous studies.
We further discuss the effect of stellar spectrum in \refsec{sec:temp}.


The high mass-loss rate due to the FUV photoelectric heating can evaporate the entire planets (see \refsec{sec:desert}).



\subsection{Metallicity dependence}\label{sec:metal}
To investigate the origin of observed sub-Jupiter desert, we study the metallicity dependence of mass-loss rates.
We run a set of simulations with gas metallicities $Z=0.05Z_{\odot}, 0.1Z_{\odot}, 0.2Z_{\odot}$.
These runs are aimed at studying the metallicity (dust-to-gas mass ratio) dependence of photo-evaporation rate driven by FUV photoelectric heating.

The atmospheric structures for the runs
with $Z=0.05Z_{\odot}, 0.1Z_{\odot}, 0.2Z_{\odot}$ are compared
in \reffig{fig:metallicity_structure}.
Since the photoelectric heating rate scales with $Z$ (Equation~[\ref{eq:pe_rate}]), the gas temperature is higher in the metal-rich 
atmosphere than in the metal-poor case.

\begin{figure}[H]
\includegraphics[width=8cm]{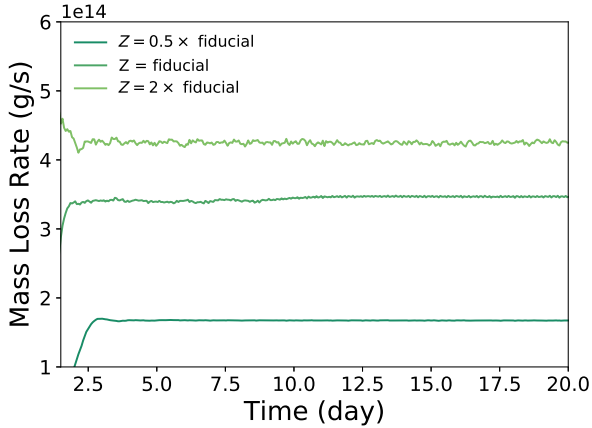}
\caption{Metallicity dependence of the mass loss rate. We calculate the mass-loss from the region defined by $|z|=2.9R_p, 0<R<2.9R_p$.}
\label{fig:metallicity_massloss}
\end{figure}

\reffig{fig:metallicity_massloss} compares $\dot{M}$ for different gas 
metallicities.
The mass-loss rate is proportional to the product of the geometrical cross section of the planetary atmosphere and the launch velocity of photo-evaporative flows as $\dot{M} \propto \pi R^2 \rho v$. 
For higher metallicities, the base radius and the velocity become larger owing to the higher opacity and the larger temperature gradient. The resulting mass-loss rates are correspondingly larger for metal-rich planets.
Our simulations suggest that the metallicity dependence of the base radius, where the photo-evaporative flows are launched, to scale approximately as $R_{\mathrm{base}}\propto Z^{0.2}$. The base density and the launch velocity scale as $\rho_{\mathrm{base}}\propto Z^{-0.5}$ and the base velocity $v_{\mathrm{base}}\propto Z^{0.8}$. 
With all these dependencies, the mass-loss rate is expected to scale as $\dot{M}\propto Z^{0.7}$ in the cases with $0.05Z_{\odot}<Z<0.2Z_{\odot}$ 
This trend results from the increasing heating rates for higher metallicity. We note that the rates of the dominant cooling are independent of the metallicity (see \refsec{sec:Fid}). The metallicity dependence suggests that it is more likely to encounter metal-poor hot Jupiters than metal-rich hot Jupiters. It contradicts higher occurrence of hot Jupiters in metal-rich systems. This may indicate that the higher occurrence originates from formation process of hot Jupiters.

\subsection{Stellar spectrum and the effective temperature}\label{sec:temp}
Atmospheric escape from planets around sun-like stars or 
around pre-main-sequence stars with low effective temperatures has been studied extensively in the literature.
Since the FUV emissivity depends strongly on the stellar effective temperature, it is important to examine cases with different stellar temperatures (spectral types). To this end, we perform additional simulations with varying the FUV flux of the host star.
We note that the EUV flux does {\it not} strongly depend 
on the stellar effective temperature \citep{Fossati_2018}. 
It is thus reasonable that we set the EUV flux to be the same as our fiducial model but vary only the FUV flux. This also makes our comparison simple and easy to interpret.
In practice, we run simulations with 1/10, 1/100, 1/1000, 1/100000 of the fiducial FUV flux. 
We fix the stellar mass and radius to the solar values.
We also run simulations with only EUV radiation. In the EUV only case, we assume the fiducial EUV flux. The results of these test runs are used to calculate and compare the mass loss caused by the EUV radiation.

\begin{figure}[H]
    \centering
    \includegraphics[width=8cm]{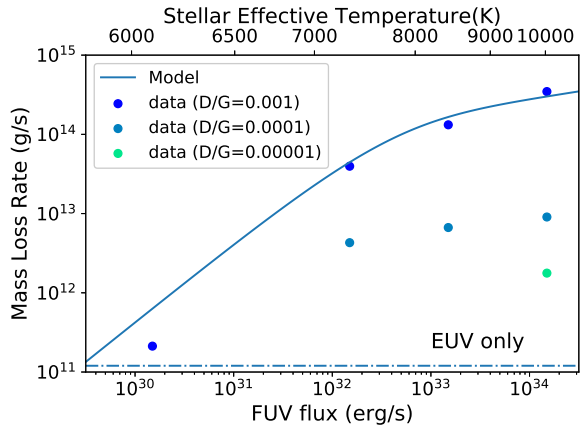}
    \caption{The flux dependence of the mass-loss rate. The points represent the simulation results and the solid curve represents our analytical model prediction (Equation~\ref{eq:massloss}). The dashed curve shows the rate of mass loss caused only by the EUV radiation. We also show the mass-loss rates for different dust-to-gas mass ratios. Our fiducial case is with D/G=0.001.  
    We only plot the cases in which the FUV dominates the escape. For even lower flux and/or lower D/G than plotted here, EUV dominates to drive the escape. 
    }
    \label{fig:FUV_flux}
\end{figure}

The resulting mass-loss rates are shown in \reffig{fig:FUV_flux}. 
For planets around hot stars ($>6000$\,K), FUV can drive the atmospheric escape, whereas EUV is important for cooler stars with $T < 6000$\,K).
We find that the mass-loss rate can be accurately estimated by considering the net heating of the atmosphere by FUV radiation as
\begin{eqnarray}
\dot{M} &=&  \epsilon\, \frac{\pi R_{p}^{3}F_{\mathrm{UV}}(\mathrm{1AU})}{GM_{p}}\left(\frac{a_{p}}{\mathrm{1AU}}\right)^{-2} \\ 
&=& 1.5\times10^{12} \,\epsilon\, \left(\frac{F_\mathrm{FUV}}{10^{30}\mathrm{erg/s}}\right)\nonumber \\
&\times&\left(\frac{R_p}{R_j}\right)^3\left(\frac{M_p}{M_j}\right)^{-1}\left(\frac{a_p}{0.045\mathrm{AU}}\right)^{-2}\,\mathrm{g/s}
\label{eq:massloss}
\end{eqnarray}
where $\epsilon$ is the heating efficiency. We adopt the FUV heating efficiency in Equation (\ref{eq:pe_rate}), and assume electron density $n_e = 10^9\ \mathrm{cm}^{-3}$ and gas temperature $T=10^4$ K. The above formula reproduces the results of our FUV + EUV runs well, as shown in \reffig{fig:FUV_flux}.
Note that the mass-loss rates in cases with EUV radiation only can be estimated as in \citet{Watson_1981,LecavelierDesEtangs_2007,Erkaev_2007} by similarly considering the heating effect with energy-limited approximation. 
 The photoelectric heating effect further increases the radius of the planet and also the electron density, which boosts the overall heating rate in a nonlinear manner.

The effect of EUV radiation is stronger than FUV radiation for stars cooler than the sun. Since the FUV photoelectric heating efficiency is much smaller than that of EUV photoionization heating \citep{Murray-Clay_2009},
close-in planets around cool stars lose the atmosphere through the EUV heating. 
However, around hot stars that emit stronger FUV radiation than the sun by several magnitudes, planets can lose their atmospheres through the FUV photoelectric heating. 
The resulting high mass-loss rate may cause apparent paucity of Hot Jupiters around hot stars.

We further quantify the FUV photoelectric heating effect with low dust-to-gas mass ratios for various stellar effective temperatures. As discussed in \refsec{sec:Fid}, metal line cooling is not dominant and the dust-to-gas ratio is a critical quantity in the present study. By varying the dust-to-gas mass ratio (D/G), we have found that the lower limit required for driving atmospheric escape by the FUV heating is D/G $\sim 10^{-5}$ in the case of 10000K host star. Below the limit, the mass-loss rate is equal to that of the EUV-only case. For host stars cooler than 10000K, the lower limit is D/G $\sim 10^{-4}$. Obviously, the lower limit depends on the FUV flux because the ratio of FUV/EUV flux also varies. We also find that the inner boundary $r = 0.85R_p $ is located at a sufficiently deeper interior than the region where the FUV radiation reaches. 
 
\section{Discussion}\label{sec:Discussion}
\subsection{Stellar metallicity dependence}\label{sec:stellar_z}
\begin{figure}[h]
    \centering
    \includegraphics[width = 8cm]{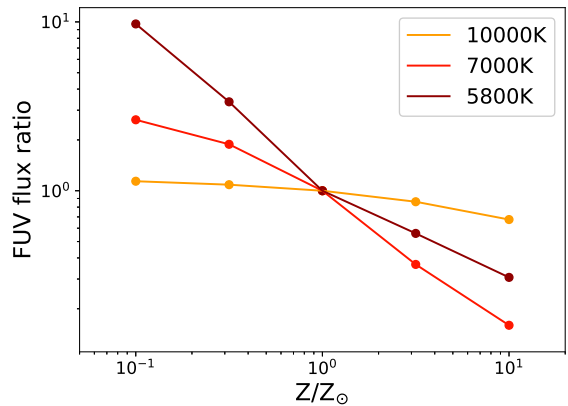}
    \caption{Stellar FUV flux and the metallicity dependence. The FUV flux ratios to that of  the $Z=Z_{\odot}$ case are shown. We calculate the FUV flux using the spectral template of \citet{Husser_2013}. 
    }
    \label{fig:FUV_metal}
\end{figure}

We have shown that the atmospheric mass-loss rate depends on the
metallicity of the planet's atmosphere. The {\it stellar} metallicity is also likely
to affect the mass-loss rate through the increased/decreased FUV flux.
We first show the metallicity dependence of the stellar FUV flux in \reffig{fig:FUV_metal}. 
We calculate the FUV flux using the theoretical stellar spectra of \citet{Husser_2013}. 
The FUV flux of metal-poor stars is larger than those of meta-rich stars
owing to the smaller opacity of the stellar atmosphere, and the metallicity effect is actually stronger for cooler stars. 

We can estimate the net metallicity dependence by incorporating both the variations in stellar FUV flux and the FUV heating rate in the planetary atmosphere. 
We assume that the metallicity of the planetary atmosphere is the same as the stellar one. Interestingly, the two effects on the mass-loss rate almost cancel out 
for cool stars with $T_{\rm eff} = 5800, 7000$\,K. The FUV flux decreases as the stellar metallicity increases, but the FUV heating efficiency increases.
As we discussed in \refsec{sec:temp}, around hot stars ($>6000$ K), FUV heating can drive atmospheric escape if the dust-to-gas mass ratio is larger than $10^{-4}$. The metallicity dependence of FUV flux is stronger in the cooler star. Thus, FUV heating can be important even in metal-poor systems with low dust-to-gas mass ratio except for very hot stars.

Overall, not only the planetary atmosphere but also the host star's metallicity critically determine the efficiency of the atmospheric escape. The net effect can even appear opposite to the case where only the planet's metallicity is considered.  


\subsection{Dust/PAHs destruction}  \label{sec:dustpah}
We have implemented photoelectric heating by small graphite grains and PAHs \citep{Bakes_1994}. 
In our simulations, the gas temperature reaches $\sim10^4$\,K, above the sublimation temperature of graphite of $\sim2200$\,K at $n_{\mathrm{H}}\sim10^{13}\ \mathrm{cm}^{-3}$\citep{Baskin_2018}. Thus the graphite grains in the hot atmosphere may sublimate, and the photoelectric heating can become inefficient. However, about a half of the photoelectric heating rate is contributed by PAH-like small grains with $N_C<1500$ carbon atoms. Hence we expect, very roughly, that the sublimation of graphite can reduce the resulting mass-loss rate by a factor of 2. 

In the photo-evaporative flows where the temperatures is $\sim10^{4}$K, PAH carbon atoms can be lost by collisions with helium \citep{Micelotta_2010b}. The PAH lifetime is estimated as
\begin{equation}
    \tau_0 = \frac{N_C}{R_T} \sim 10^{6-10} \mathrm{s}
\end{equation}
where $R_T$ is the rate coefficient for collisional destruction. 
This lifetime should be compared with the hydrodynamical timescale, the sound crossing time 
of $\tau=R_p/c_s\sim10^{4}$\,s. 
Clearly, destruction of PAHs is unimportant 
in the escaping atmosphere of the hot Jupiters we study here.

\subsection{Other heating processes}\label{sec:other}
FUV radiation is absorbed through the hydrogen Balmer series transitions and can drive the so-called Balmer escape. \citet{Munoz_2019} suggest that $\dot{M}$ of the Balmer escape is similar to what we find in our simulations.
Rapid mass-loss has been observed around ultra-hot Jupiter, KELT-9 b \citep{Wyttenbach_2020}, which 
is undergoing mass-loss consistent with Balmer escape and with our model. Since Balmer escape does not depend on the metallicity of the atmosphere, the mass loss rate is expected to depend largely on the host star's FUV flux.
It would be interesting to understand the relative importance and the combined effect of these two processes as a function of metallicity.

X-ray heating is also effective to drive atmospheric escape from planets around active stars
\citep{Owen_2012}. Young stars, e.g. pre-main-sequence stars, emit strong X-ray radiation to drive atmospheric escape. X-rays can penetrate deep into the interior of the planet's atmosphere, and increase the electron density and FUV heating rate in a coupled manner.
The "boosting" effect is observed in simulations of protoplanetary disk photo-evaporation \citep{Gorti_2009, Nakatani_2018b}. 
We shall investigate the effects of these heating processes in our future work.

\subsection{Sub-Jovian desert}\label{sec:desert}
Our simulations suggest that a close-in giant planet can lose a large fraction of its atmosphere in  $t_{\mathrm{esc}}=M_p/\dot{M} \sim 10^{8-9}$ yrs. 
In our fiducial model, the planetary atmosphere can actually be lost in $6\times10^{7}$ yrs, which is shorter than the lifetime of a hot A-star. Therefore,
the so-called sub-Jovian desert may be explained by the FUV-driven atmospheric escape. 

The photoelectric heating efficiency depends on the amount of dust grains in the atmosphere, and thus we expect the photo-evaporation rate scales with the atmosphere's metallicity.
We have run a set of simulations and have shown that the mass-loss rate is systematically larger for metal-rich planets than for metal-poor planets (\refsec{sec:metal}). 
Interestingly, the trend appears opposite to the observation that the sub-Jovian desert is more prominent for planets around metal-poor stars.
We argue that the net dependence may be more complicated because the stellar SED also depends on the host star's metallicity.
In fact, for cool stars, the metallicity dependence nearly cancels out (\refsec{sec:stellar_z}) and thus the atmospheric escape rate does not sensitively depend on the metallicity.

The abundance of sub-Jupiters around metal-rich stars may be explained by the planet-metallicity correlation \citep{Fischer_2005}, because the mass-loss rate of planets around cool stars does not strongly depend on the metallicity (\refsec{sec:stellar_z}). Our results suggest that the sub-Jovian desert may appear different for hot star systems may and for cool star systems. Close-in planets around metal-rich, hot stars lose the atmosphere quickly, which can shape the sub-Jovian desert. Further studies on planet population synthesis that incorporate the atmospheric escape will reveal the origin of the observed sub-Jupiter desert. 

\section{Conclusion}\label{sec:summary}
Atmospheric escape is one of the key physical processes in the study of exoplanet populations.
Previous theoretical studies have clarified the effect of EUV photoionization heating, and our present study suggests that FUV photoelectric heating can also be important in close-in planets if the planetary atmosphere contains a significant amount of dust grains. The FUV heating effect can be even dominant for planets around hot, A-stars. 
Our radiation hydrodynamics simulations show that the FUV radiation can raise the gas temperature to 10000 K. The effective gas heating drives atmospheric escape so strongly that a large fraction of the atmosphere can be lost in about 100 million years. The FUV driven mass loss rate $\dot{M}$ is larger for metal-rich planets.
The outflows driven by the FUV heating can transport small dust grains from the interior to the upper atmosphere. For FUV heating to be strong to drive escape, the dust-to-gas mass ratio should be $D/G>10^{-5}$ for host stars ($\sim10000$K).

The observed metallicity dependence of the sub-Jovian desert, i.e. the apparent paucity of close-in planets around metal-poor stars, is intriguing.
Our results may explain the observation {\it if} the metallicity dependence of the host star's FUV flux is properly taken into account.  Metal-rich stars with weaker FUV radiation drive atmospheric escape less effectively. We argue that considering both
the planet's and the host star's metallicities is important to understand the statistics of close-in planets. Future observations of populations of close-in planets will clarify the metallicity dependence, both of planets and of host stars, and then will help us with understanding the physical mechanism of atmospheric escape.

\acknowledgements
We thank Kazumasa Ono for discussion on planetary atmosphere. We are also grateful to the anonymous referee for insightful comments.
HM has been supported by International Graduate Program for Excellence in Earth-Space Science (IGPEES) of the University of Tokyo.
RN is supported by Grant-in-Aid for Research Activity Start-up (19K23469) and the Special Postdoctoral Researcher program at RIKEN.
The numerical computations were carried out on Cray XC50 at Center for Computational Astrophysics, National Astronomical Observatory of Japan.

\bibliographystyle{aasjournal}
\bibliography{atmospheric_escape}



\end{document}